\documentclass[12pt,preprint]{aastex}

\slugcomment{Submitted to ApJ Lett May 7, 2009, Resubmitted May 21}

\shorttitle{Follow-up of HAT-P-11b}
\shortauthors{Dittmann et al.}

\begin{document}

%% LaTeX will automatically break titles if they run longer than
%% one line. However, you may use \\ to force a line break if
%% you desire.

\title{Follow-up Observations of the Neptune Mass Transiting Extrasolar Planet HAT-P-11b}

%% Use \author, \affil, and the \and command to format
%% author and affiliation information.
%% Note that \email has replaced the old \authoremail command
%% from AASTeX v4.0. You can use \email to mark an email address
%% anywhere in the paper, not just in the front matter.
%% As in the title, use \\ to force line breaks.

\author{Jason A. Dittmann, Laird M. Close, Elizabeth M. Green, Louis J. Scuderi, and Jared R. Males}
\affil{Steward Observatory, University of Arizona, Tucson, AZ 85721}

%\author{S. Djorgovski\altaffilmark{1,2,3} and Ivan R. King\altaffilmark{1}}
%\affil{Astronomy Department, University of California,
%    Berkeley, CA 94720}%

%\author{C. D. Biemesderfer\altaffilmark{4,5}}
%\affil{National Optical Astronomy Observatories, Tucson, AZ 85719}
%\email{aastex-help@aas.org}

%\and

%\author{R. J. Hanisch\altaffilmark{5}}
%\affil{Space Telescope Science Institute, Baltimore, MD 21218}

%% Notice that each of these authors has alternate affiliations, which
%% are identified by the \altaffilmark after each name.  Specify alternate
%% affiliation information with \altaffiltext, with one command per each
%% affiliation.

%\altaffiltext{1}{Visiting Astronomer, Cerro Tololo Inter-American Observatory.
%CTIO is operated by AURA, Inc.\ under contract to the National Science
%Foundation.}
%\altaffiltext{2}{Society of Fellows, Harvard University.}
%\altaffiltext{3}{present address: Center for Astrophysics,%
%    60 Garden Street, Cambridge, MA 02138}
%\altaffiltext{4}{Visiting Programmer, Space Telescope Science Institute}
%\altaffiltext{5}{Patron, Alonso's Bar and Grill}

%% Mark off your abstract in the ``abstract'' environment. In the manuscript
%% style, abstract will output a Received/Accepted line after the
%% title and affiliation information. No date will appear since the author
%% does not have this information. The dates will be filled in by the
%% editorial office after submission.

\begin{abstract}

We have confirmed the existence of the transiting super Neptune
extrasolar planet HAT-P-11b. On May 1, 2009 UT the transit of
HAT-P-11b was detected at the University of Arizona's 1.55m Kuiper
Telescope with 1.7 millimag rms accuracy. We find a central transit
time of $T_c = 2454952.92534\pm0.00060$ BJD; this transit occurred
$80\pm73$ seconds sooner than previous measurements (71 orbits in the
past) would have predicted. Hence, our transit timing rules out the
presence of any large ($>200$s) deviations from the ephemeris of Bakos
et al. (2009). We obtain a slightly more accurate period of
$P=4.8878045\pm0.0000043$ days. We measure a slightly larger planetary
radius of $R_p=0.452\pm0.020 R_{J}$ ($5.07\pm0.22R_\earth$) compared
to Bakos and co-workers' value of $0.422\pm0.014 R_{J}$ ($4.73\pm0.16
R_\earth$). Our values confirm that HAT-P-11b is very similar to GJ
436b (the only other known transiting super Neptune) in radius and
other bulk properties.

\end{abstract}

%% Keywords should appear after the \end{abstract} command. The uncommented
%% example has been keyed in ApJ style. See the instructions to authors
%% for the journal to which you are submitting your paper to determine
%% what keyword punctuation is appropriate.


%% From the front matter, we move on to the body of the paper.
%% In the first two sections, notice the use of the natbib \cite
%% and \citet commands to identify citations.  The citations are
%% tied to the reference list via symbolic KEYs. The KEY corresponds
%% to the KEY in the \bibitem in the reference list below. We have
%% chosen the first three characters of the first author's name plus
%% the last two numeral of the year of publication as our KEY for
%% each reference.

%% Authors who wish to have the most important objects in their paper
%% linked in the electronic edition to a data center may do so by tagging
%% their objects with \objectname{} or \object{}.  Each macro takes the
%% object name as its required argument. The optional, square-bracket 
%% argument should be used in cases where the data center identification
%% differs from what is to be printed in the paper.  The text appearing 
%% in curly braces is what will appear in print in the published paper. 
%% If the object name is recognized by the data centers, it will be linked
%% in the electronic edition to the object data available at the data centers  
%%
%% Note that for sources with brackets in their names, e.g. [WEG2004] 14h-090,
%% the brackets must be escaped with backslashes when used in the first
%% square-bracket argument, for instance, \object[\[WEG2004\] 14h-090]{90}).
%%  Otherwise, LaTeX will issue an error. 

\section{Introduction}

The transit of an extrasolar planet across the face of its host star
allows direct measurement of the bulk properties of the planet. In
particular, the transit allows accurate determination of the planet's
radius (see for example Charbonneau et al. 2006 and references
within). When these radii are combined with radial velocity (RV)
measurements of masses then 
densities of the transiting extrasolar planets can be
calculated. Knowledge of the heating from the star allows estimates of
the temperatures of the irradiated planets. Models of the bulk properties of these planets can be
compared to observation (see for example Baraffe et al. 2008; Fortney
et al. 2007; Burrows et al.  2007; Seager et al. 2007).

While some 59\footnote{http://exoplanet.eu} transiting extrasolar planets are now known,
only 3 have masses less than 10\% of Jupiter (closer to Neptune in
mass). The first transiting super Neptune was GJ 436b which was
discovered around a M2.5 star at 0.028 AU by an RV survey of Butler et al. (2004). They found it had a
mass of $\sim$21 $M_\earth$. Follow-up photometric observations of GJ 436b then discovered it to be a $\sim$7-8 millimag ($\sim$0.7\%) transiting planet (Gillon et al. 2007). Further follow-up measurements find a
radius of $\sim4.2-4.9R_\earth$ (Torres, Winn \& Holman 2008; Bean et
al. 2008; respectively) and a density of $\rho=1.69^{+0.14}_{-0.12} g cm^{-3}$ (Torres 2008). Through these transit observations and
modeling by Baraffe et al. (2008) it has been determined that GJ 436b is
mainly composed of metals with only a small H/He envelope.

The second transiting Neptune discovered was HAT-P-11b, in orbit
around HAT-P-11 (2MASS 19505021+4804508) a K4 (V=9.6 mag) metal rich
star. This planet was discovered by the HATNet array of small 0.11m
telescopes on Mt. Hopkins in Arizona (Bakos et al. 2009). At just 4.2
millimag (mmag) HAT-P-11b was the smallest planet discovered by the
transit method. Moreover, Bakos et al. (2009) argue that compared to
recent measurements of the radius of GJ 436b, HAT-P-11b was at the time the smallest transiting extrasolar planet known. However, the
CoRoT team has announced the discovery of COROT-7b which is smaller
still with just 1.7 $R_\earth$ (Rouan et al. 2009). In any case, transiting objects of less than 0.1 $M_{J}$
number no more than three today and start to probe densities and
masses closer to terrestrial --in contrast to lower density gas giants
composed mainly of a large H/He envelope.

In the case of HAT-P-11b, detailed RV measurements by Bakos et
al. (2009) show a linear drift ($0.0297\pm0.0050$ m/s/day) in the RV
residual of HAT-P-11. This drift is possibly due to the pull of an
additional unseen planet in the system (Bakos et al. 2009). In
addition, Bakos et al. (2009) determine a non-zero (0.198$\pm$0.046)
eccentricity which might be maintained by interactions with another
planet. Both observations hint at the presence of another outer planet
``HAT-P-11c'' in the system. Indeed, most systems with super Neptunes
are multiple planet systems (Bakos et al. 2009 and references within). However, it is worth noting that to date
no transiting planet is known to be a member of a multiple planet
system. Hence, detection of multiple transiting planet system would be
very interesting. Continued RV monitoring of this system may directly
detect curvature in the RV residuals due to this outer planet. Another
way to directly detect the presence of a possible ``HAT-P-11c'' would be a
sensitive search for transit timing variations over a series of
HAT-P-11b transits. A search to bound the magnitude of such timing
variations motivated this paper.

\section{Observations \& Reductions}

Data was taken at the University of Arizona's 61-inch (1.55m) Kuiper
telescope on Mt.\ Bigelow, Arizona on 1 May, 2009 UT with the Mont4k
CCD, binned 3x3 to 0.43$^{\prime\prime}$/pix.  Observing a single
transit of HAT-P-11b at high S/N is fairly challenging, since the
transit depth is just $\sim$ 4.3 mmag, HAT-P-11 itself is a very
bright $V$ = 9.6 magnitude K4 star with a nearby faint (likely
background) star, the surrounding ``Kepler field'' is somewhat crowded,
and all of the potential photometric reference stars are several
magnitudes fainter than HAT-P-11.  Defocusing was not possible, due to
the crowded field and the difficulty of maintaining a consistent focus
offset with this telescope/instrument combination.  In order to take
long enough exposures to sufficiently average out atmospheric
scintillation noise, while still avoiding saturation of the ccd, we
used a medium bandwidth Stromgren b filter ($\Delta\lambda=$18.0 nm).  The Mont4k filter holder and filter sensor
position were specifically modified to accommodate the thicker
Stromgren filter for our observations.

The conditions were photometric with light wind and no moon throughout
the observational period.  In total 448 images were obtained with $<2$
pixels of wander, due to excellent autoguiding. Integration times of
20~s were used at the start when the target was still at relatively
high airmass. Just prior to the start of the transit (after the $109^{th}$ image), the exposure
times were reset to 17~s, giving a sampling time of 28.3~s for the
remainder of the observations. The relatively fast overhead time is
achieved mainly by binning 3x3 and skipping the flushing of the ccd
after each readout and before the subsequent exposure in the sequence,
but is also a product of the Mont4k's design, which includes two
amplifiers and pre-amplifiers.

The images were bias-subtracted, flat-fielded, and bad pixel-cleaned
in the usual manner.  Aperture photometry and sky subtraction was
performed for the target star and three reference stars using the
aperture photometry task PHOT in the IRAF DAOPHOT
package\footnote{IRAF is distributed by the National Optical Astronomy
Observatories, which are operated by the Association of Universities
for Research in Astronomy, Inc., under cooperative agreement with the National Science Foundation.}. An aperture radius of
4.3$^{\prime\prime}$ (10.0 pixels) was adopted, as it produced the
smallest scatter in the light curve and also eliminated contamination
from HAT-P-11's nearby companion, which is 5.4 magnitudes fainter in b
and 8.9$^{\prime\prime}$ to the NNE (at a position angle of
9$^\circ$). The reference stars are 2.3 to 2.65 magnitudes fainter
than HAT-P-11, and were chosen to be distributed as uniformly as
possible about the target star on the sky (they formed a triangle
around HAT-P-11 at distances of 89, 208, $345\arcsec$ from
HAT-P-11). The reference stars were
normalized to unity and then weighted according to their average
fluxes.

We applied no sigma clipping rejection to the reference stars or
HAT-P-11b --all datapoints were used in the analysis. The final light
curve for HAT-P-11 was normalized by division of the weighted average
of the three reference stars. The residual light curve in Fig. 1
(bottom left) has a photometric RMS range of 1.7 mmag rms and a time
sampling of 29 seconds on average. This is very typical of the
relative photometric precision achieved with the Mont4k on the 61-inch (Kuiper)
telescope for high S/N images (Randall et al. (2007); Dittmann et
al. 2009).

\section{Analysis}
\subsection{A Search for Transit Timing Variations}

      The planetary transit light curves were fit using the $\chi^{2}$
method prescribed by Mandel and Agol (2002). The transit HAT-P-11b
parameters used in the fit were those in Table 1 measured by Bakos et
al. (2009). The correct linear and quadratic limb darkening parameters
for our b filter were taken from Claret (2000). In order to detect any
transit timing variations the only parameter that was allowed to vary
in the fit was the central time of of the transit, $T_c$ . The time of
the center of this transit, ($T_c$), is shown near the bottom of Table 1. We note that
the purpose of this section of the paper was not to re-derive all the
parameters of the transit but to understand if our transit is
consistent with the period of Bakos et al. (2009).

      To measure $T_c$ we minimized $\chi^{2}$ to find a $T_c =
      2454952.92534\pm0.00060$ BJD value (the flux uncertainty for
      each datapoint in the $\chi^{2}$ fit was calculated from the
      propagation of the photometric errors determined with the PHOT
      task). The $\pm0.00060$ day $1\sigma$ uncertainty was
      estimated by Monte-Carlo simulations of 1000 simulated datasets
      with the same 1.7 mmag rms scatter of the original data (see
      Fig. 2 left).

\subsection{Determination of a new Planetary Radius Value}

    To try and understand if our transit data suggest a different
planetary radius for HAT-P-11 we repeated the fit in the last section using the $\chi^{2}$
method prescribed by Mandel and Agol (2002) but this time allowed the
planetary radius $R_p$ to vary along with $T_c$. In Fig. 1 (right) we see the result of our
fit (solid red line) and the residuals of this fit below.

      To measure $R_p/R_*$ we further minimized the reduced $\chi_{\nu}^{2}$
      to 1.06 with simultaneous fits of $R_p/R_*=0.0621\pm0.0011$ and $T_c =
      2454952.92534$.  The $\pm0.0011$ $1\sigma$
      uncertainty in $R_p/R_*$ was estimated by Monte-Carlo
      simulations of 1000 fake datasets with the same 1.7 mmag rms
      scatter as the original data (see Fig. 2 right).

\section{Discussion}

\subsection{Is the Timing of the Transits Changing?}

A key goal of this paper is to compare our measured May 1, 2009 UT
$T_c$ to that predicted from the previously measured
values. Projecting the P=$4.8878162\pm0.0000071$ day period of Bakos
et al. (2009) forward from their most accurate $T_{c1}
=2454605.89132\pm0.00032$ transit suggests that our May 1 transit
($n$=71 periods later) occurred $\Delta T=80\pm73$ seconds sooner
(where $\sigma_{\Delta T}$ was calculated by
$\surd(\sigma_{Tc1}^2+n\sigma_{P}^2+\sigma_{Tc2}^2)$) than our
observed $T_{c2} = 2454952.92534\pm0.00060$ BJD was predicted to
be. However, the significance of this disagreement is small. Indeed
there is $\sim$60\% probability that our observations are fully
consistent with the timing measurements (and uncertainties) of Bakos
et al. (2009). Certainly, we can rule out large $>200$ second timing errors at the $\sim3 \sigma$ level. 

With the addition of our new $T_c$ values to the two previous values
we derive a new P=$4.8878045\pm0.0000043$ day value for the period of
HAT-P-11b (based on a sigma weighted average; see Table 1). This new value
is slightly shorter than the P=$4.8878162\pm0.0000071$ day period of
Bakos et al. (2009). However, it will require future observations to
determine if this new period will better predict future transit
times. It is entirely possible that all of our transit timing values
are consistent with predictions from Bakos et al.'s ephemeris within
measurement errors. Moreover, one can calculate that the period
estimate between the $T_c$ measurements over the first 284 orbits of
Bakos et al. (2009) and over the last 71 orbits has only changed by
$0.177\pm1.009$ seconds. Hence, there is no significant evidence, with
the data in hand, that Hat-P-11b's period has changed over the last
0.95 year compared to the previous 3.79 years.

In general, it is difficult with a single additional transit to
confidently determine if the discovery period of Bakos et al. is
changing with time. However, our $T_{c2}$ datapoint 71 orbits later
yields a more accurate period of P=$4.8878045\pm0.0000043$ day. This
new period with its lower error can be used to construct a standard
O-C (observed minus calculated time-of-transit) diagram which will
start showing ``non-linear'' trends in the future O-C residuals if the
period is truly changing. 

\subsection{What is the Radius of HAT-P-11b?}

  Our observations of the depth of the transit finds a deeper transit
  and a $R_p/R_*=0.0621\pm0.0011$ compared to $0.0576\pm0.0009$ of
  Bakos et al. (2009). We find that $R_p/R_*$ is $0.0045\pm0.0014$
  larger than that of Bakos et al. (2009). Hence, there is a significant
  probability that our transit was deeper than that of Bakos et
  al. (2009). We derive a slightly larger planetary radius of
  $R_p=0.452\pm0.020$ $R_{J}$ ($5.07\pm0.22 R_\earth$) compared to
  Bakos et al.'s values of $0.422\pm0.014 R_{J}$ ($4.73\pm0.16
  R_\earth$). Our values suggest that HAT-P-11b is very similar to GJ 436b
  in radius.

%  There is still a need for future work in the search for wider
%companions to HAT-P-11b. ({\bf add Jared's PM search here?}).

\section{Conclusions}

We confirm the existence of the transiting planet HAT-P-11b. Our
main conclusions from our 1.7 mmag rms (unbinned) transit
observations (with the University of Arizona's 1.55 Kuiper telescope) of
the May 1, 2009 UT transit are:

1. We find a central transit time of $T_c = 2454952.92534\pm0.00060$
BJD from a best fit to our data. We estimate that the transit occurred
$80\pm73$ seconds sooner  than previous (71 orbits in the past) measurements would have
predicted (Bakos et al. 2009). Our finding is consistent with the ephemeris of Bakos et al. and rules out the presence of any large timing variation.

2.  We derive a slightly larger planetary radius of
  $R_p=0.452\pm0.020$ $R_{J}$ ($5.07\pm0.22 R_\earth$) compared to
  Bakos et al.'s values of $0.422\pm0.014 R_{J}$ ($4.73\pm0.16
  R_\earth$). Our values suggest that HAT-P-11b is very close to GJ 436b
  in radius.

\acknowledgments

  We would like to thank the anonymous referee for helpful comments
leading to a better final paper. We would like to thank Greg Stafford,
Gary Rosenbaum, and the Catalina Mountain staff for modifying the
Mont4k filter box and holder so that we could use the Stromgren b
filter. We would also like to thank the Arizona NASA Space Grant
program for funding this work. LMC is supported by a NSF Career award
and the NASA Origins program.

%% To help institutions obtain information on the effectiveness of their
%% telescopes, the AAS Journals has created a group of keywords for telescope
%% facilities. A common set of keywords will make these types of searches
%% significantly easier and more accurate. In addition, they will also be
%% useful in linking papers together which utilize the same telescopes
%% within the framework of the National Virtual Observatory.
%% See the AASTeX Web site at http://www.journals.uchicago.edu/AAS/AASTeX
%% for information on obtaining the facility keywords.

%% After the acknowledgments section, use the following syntax and the
%% \facility{} macro to list the keywords of facilities used in the research
%% for the paper.  Each keyword will be checked against the master list during
%% copy editing.  Individual instruments or configurations can be provided 
%% in parentheses, after the keyword, but they will not be verified.

{\it Facilities:} \facility{Kuiper 1.55m}.

\clearpage 
{\bf References}\\

Bakos et al. 2009 arXiv 0901.0282 submitted to ApJ\\

Baraffe, I., Chabrier, G., \& Barman, T. 2008, A\&A, 482, 315\\

Bean, J. L., et al. 2008, A\&A, 486, 103949\\

Burrows, A., Hubeny, I., Budaj, J., \& Hubbard, W. B. 2007, ApJ,
661, 502\\

Butler, R.P., Wright, J.T., Marcy, G.W., Fischer, D.A., et al.. 2006 ApJ 646, 505.\\

Charbonneau, D., Brown, T., Burrows, A., Laughlin, G. 2006. Protostars and Planets V.(astro-ph/0603376v1)\\

Claret, A. 2004, A\&A, 428, 1001\\

Dittmann, J. A., Close, L.M., Green, E.M., Fenwick, M. 2009, ApJ submitted.\\

Fortney, J. J., Marley, M. S., \& Barnes, J. W. 2007, ApJ, 659, 1661\\

Mandel, K., \& Agol, E. 2002, ApJ, 580, L171\\

Neter et al. Applied Statistics, 3rd ed. Boston : Allyn and Bacon, 1982\\

Randall, S.K., Green, E.M., van Grootel, V., et al. 2007 A\&A , 476, 1317\\
Rodono, M., Lanza, A. F., Catalano, S. 1995. A\&A. 301, 75.\\

Rouan et al. 2009; CoRoT International Symposium\\

Seager, S., Kuchner, M., Hier-Majumder, C. A., \& Militzer,
B. 2007, ApJ, 669, 1279\\

Torres, G. 2007, ApJ 671, 65\\

Torres, G., Winn, J. N., Holman, M. J. 2008, ApJ, 677, 1324\\

%\bibitem[Auri\`ere(1982)]{aur82} Auri\`ere, M.  1982, \aap,
%    109, 301\\

%\bibitem[Canizares et al.(1978)]{can78} Canizares, C. R.,
%    Grindlay, J. E., Hiltner, W. A., Liller, W., \&
%    McClintock, J. E.  1978, \apj, 224, 39
%\bibitem[Djorgovski \& King(1984)]{djo84} Djorgovski, S.,
%    \& King, I. R.  1984, \apjl, 277, L49
%\bibitem[Hagiwara \& Zeppenfeld(1986)]{hag86} Hagiwara, K., \&
%    Zeppenfeld, D.  1986, Nucl.Phys., 274, 1

%\end{thebibliography}

\clearpage

%% Use the figure environment and \plotone or \plottwo to include
%% figures and captions in your electronic submission.
%% To embed the sample graphics in
%% the file, uncomment the \plotone, \plottwo, and
%% \includegraphics commands
%%
%% If you need a layout that cannot be achieved with \plotone or
%% \plottwo, you can invoke the graphicx package directly with the
%% \includegraphics command or use \plotfiddle. For more information,
%% please see the tutorial on "Using Electronic Art with AASTeX" in the
%% documentation section at the AASTeX Web site,
%% http://www.journals.uchicago.edu/AAS/AASTeX.
%%
%% The examples below also include sample markup for submission of
%% supplemental electronic materials. As always, be sure to check
%% the instructions to authors for the journal you are submitting to
%% for specific submissions guidelines as they vary from
%% journal to journal.

%% This example uses \plotone to include an EPS file scaled to
%% 80% of its natural size with \epsscale. Its caption
%% has been written to indicate that additional figure parts will be
%% available in the electronic journal.

\begin{table}
\begin{center}
\caption{Parameters of the HAT-P-11 system\label{tbl-1}}
\begin{tabular}{crrr}
\tableline\tableline
  Parameter & Value & Reference \\
  \tableline
   P (days) & 4.8878162 $\pm$ 0.0000071 & Bakos et al. (2009) \\ 
   $T_{c}$ (BJD) & 2453217.75466 $\pm$ 0.00187 & Bakos et al. (2009) \\
   $T_{c}$ (BJD) & 2454605.89132 $\pm$ 0.00032 & Bakos et al. (2009) \\ 
   b & $0.347^{+0.130}_{-0.139}$ & Bakos et al. (2009) \\ 
   $i$ (deg) & $88.5 \pm 0.6$ & Bakos et al. (2009) \\
   $R_*$ ($R_\sun$) & $0.75\pm0.02$ & Bakos et al. (2009)\\ 
   $R_{p}/R_{*}$ & 0.0576 $\pm$ 0.0009 & Bakos et al. (2009) \\ 
   $M_{p}$ ($M_{J}$) & $0.081 \pm 0.009$ & Bakos et al. (2009) \\ 
   $R_{p}$ ($R_{J}$) & $0.422 \pm 0.014$ & Bakos et al. (2009) \\ 
  \tableline
  $T_{c}$ (BJD) & 2454952.92534 $\pm$ 0.00060 & This work \\ 
  P (days) & 4.8878045\tablenotemark{a}$\pm$ 0.0000043 & This work \\ 
  $R_{p}/R_{*}$ & 0.0621 $\pm$ 0.0011 & This work \\ 
  $R_{p}$ ($R_{J}$) & $0.452 \pm 0.020$ & This work \\ 
\tableline
\end{tabular}

\tablenotetext{a}{the new period value was calculated by a sigma weighted least square of all three $T_c$ values}
%\cite{low05}, the errors are as given by \cite{low05}.}
%\tablenotetext{b}{data from \cite{low05}.} 
%\tablenotetext{c}{data from
%our WiFi ADI reduction, astrometric errors dominated by 0.5\%
%platescale errors across the field. }

\end{center}
\end{table}

\clearpage

\begin{figure}
\epsscale{1.0}
\plottwo{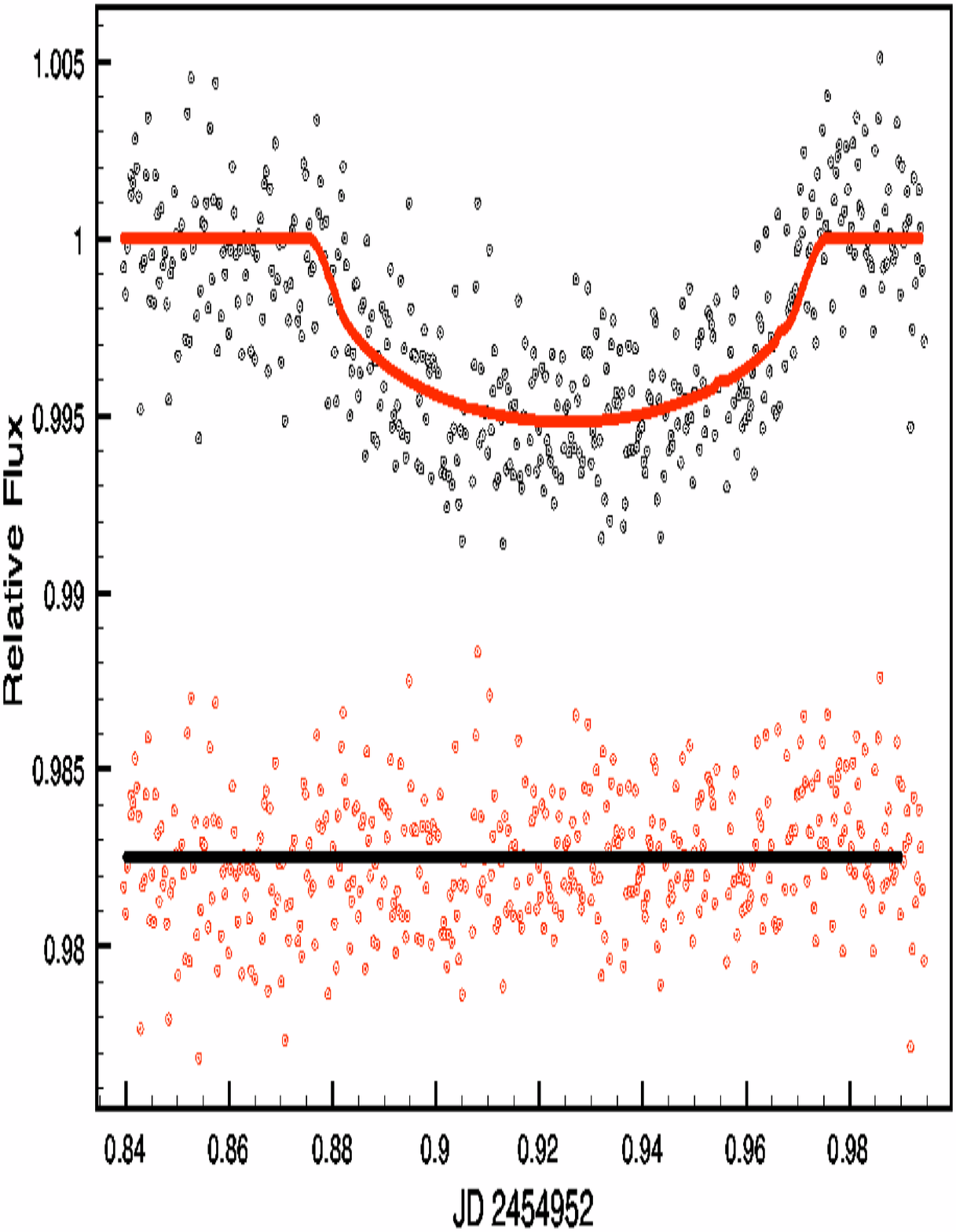}{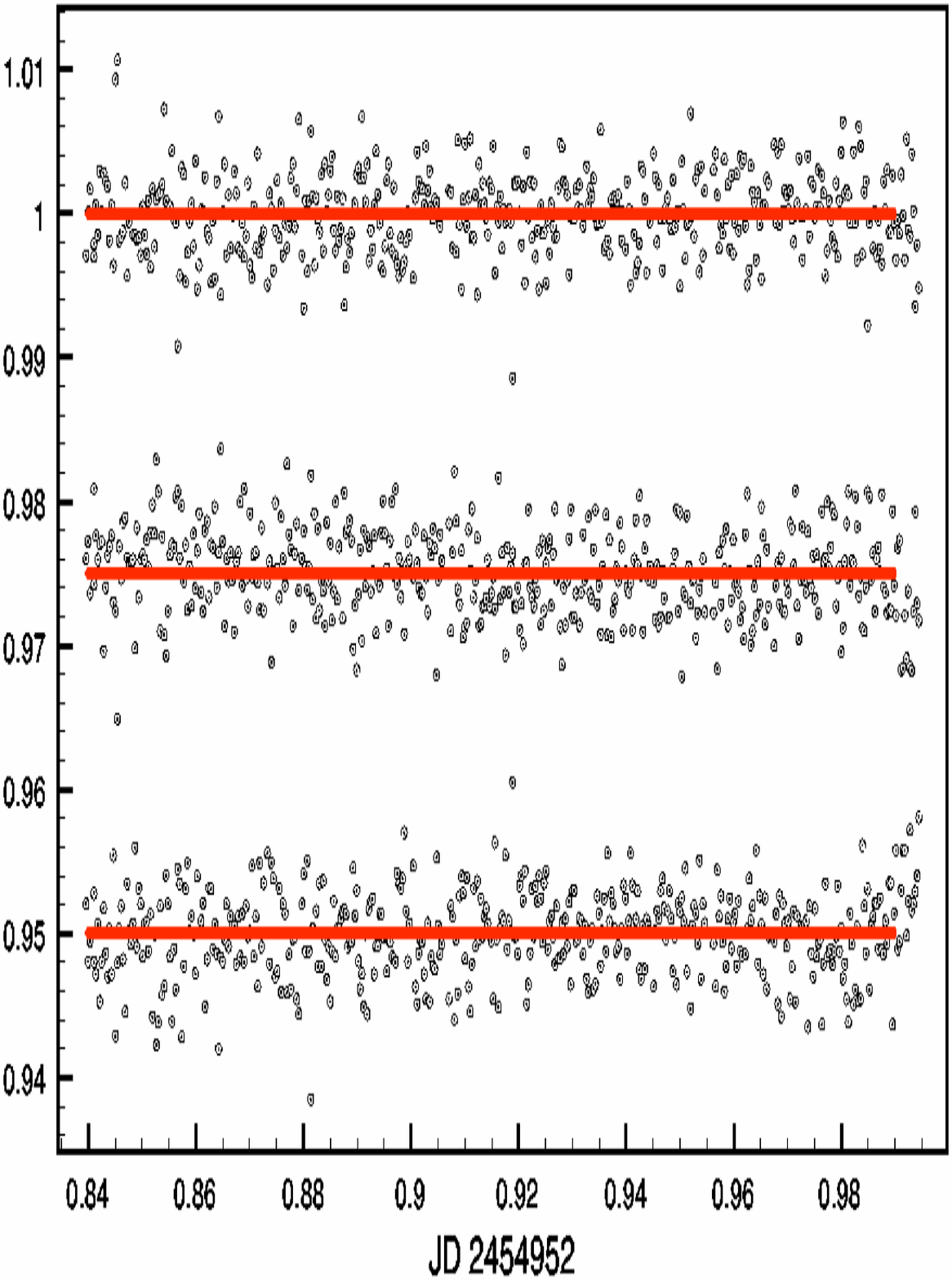}
\caption{{\bf Left:} The timeseries of HAT-P-11 during the Transit of
May 1, 2009 UT. We show our best fit (reduced $\chi_{\nu}^{2}$ =1.06)
with simultaneous fits of $R_p/R_*=0.0621\pm0.0011$ and $T_c =
2454952.92534$ ( solid red curve). The 1.7 mmag rms residuals of the
fit are shown below.  {\bf Right:} The timeseries of our three
calibrator stars (each normalized by the sum of the remaining two
calibrator stars). The excellent conditions of the night allowed for
mmag photometry in individual 17 or 20 second exposures even on
these fainter reference stars.
\label{fig1}}
\end{figure}

\clearpage

\begin{figure}
\epsscale{1.0}
\plottwo{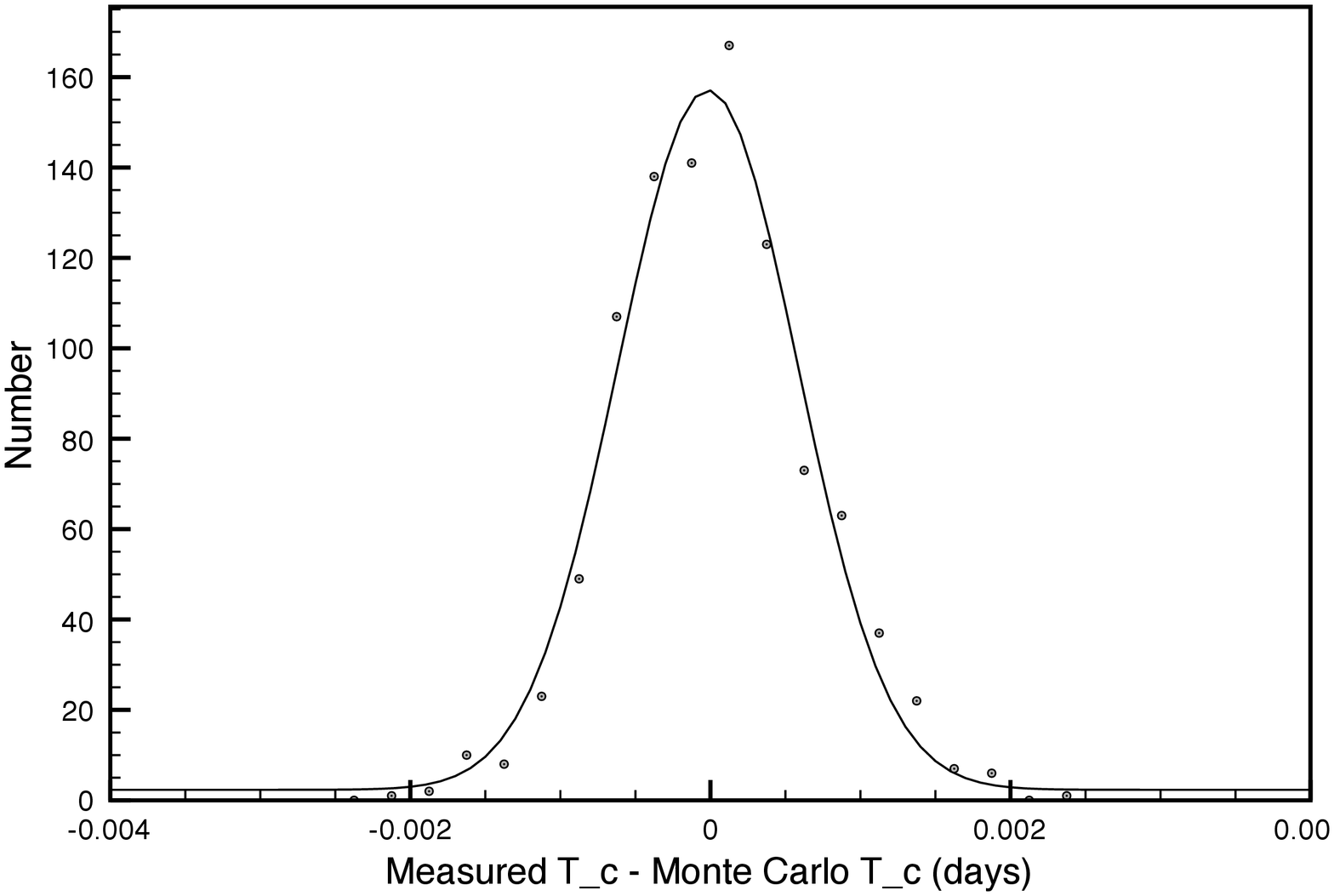}{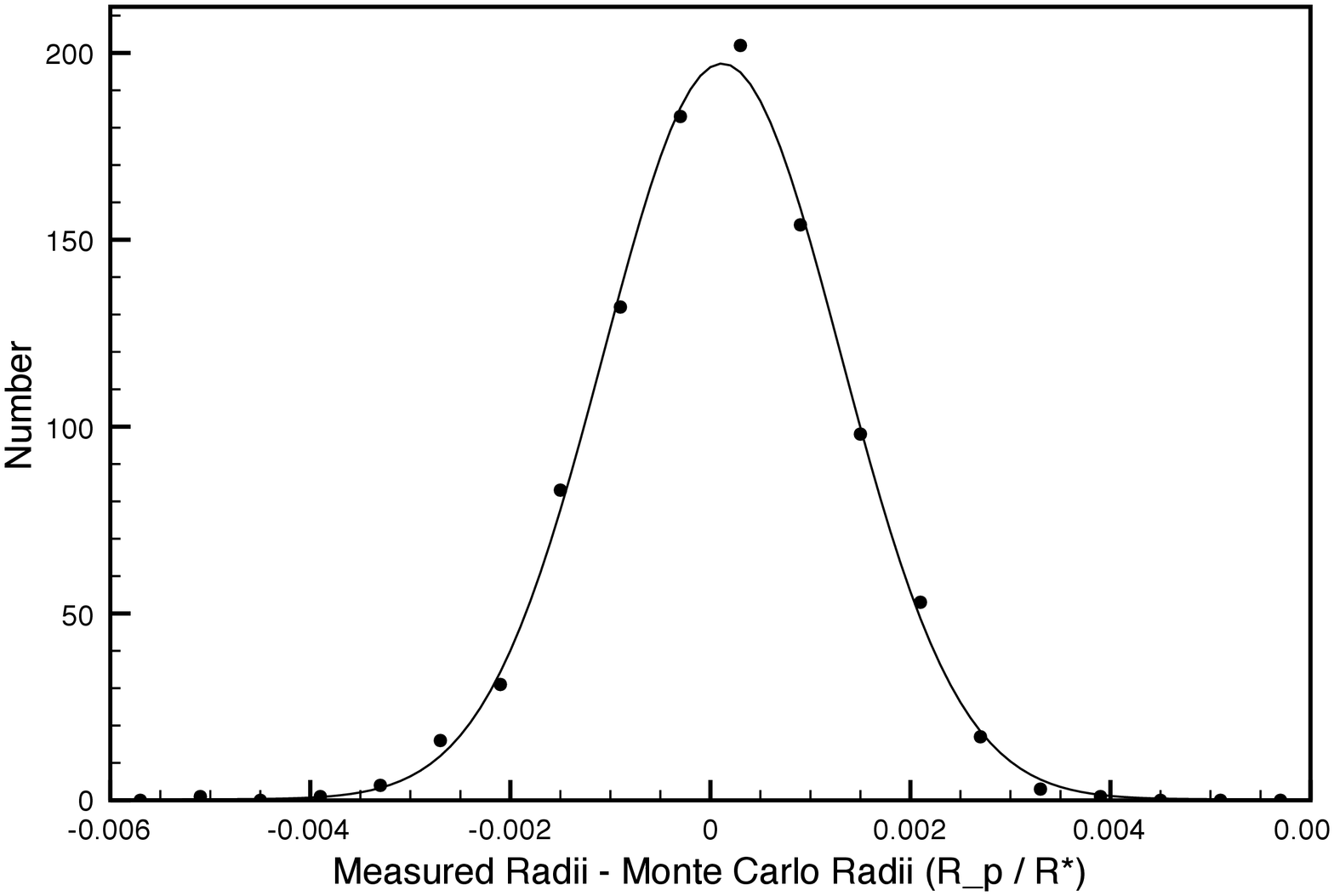}
\caption{{\bf Left:} One thousand Monte-Carlo (MC) simulations of
independent simulated datasets following our best-fit transit model
each drawn from a population of data with the same 1.7 mmag rms as our
data in Fig. 1 (left). Based these MC simulations we find with our 1.7 mmag
rms uncertainty we can constrain $T_c$ to an accuracy of $\pm0.00060$
days (or 51.84 seconds) at the 1$\sigma$ level.  {\bf Right:} Here another one
thousand MC realizations imply the uncertainty in the $R_p/R_*$ ratio
to be $\pm0.0011$ at the 1$\sigma$ level.
\label{fig2}}
\end{figure}

%% The following command ends your manuscript. LaTeX will ignore any text
%% that appears after it.

\end{document}